# Understanding Decision-Making Across the Lifespan Needs Theoretical Neuroscience

Michael B. Ryan[1], Letizia Ye[1], Anne K. Churchland[1]

[1]Department of Neurobiology, David Geffen School of Medicine, University of California, Los Angeles

**Bullets**
- Computational tools in theoretical neuroscience are mostly tested on young brains
- Aging alters cellular and circuit function, providing biological constraints for existing models
- Older brains may afford an opportunity to uncover novel neural states and dynamics

**Abstract**
Understanding how decision-making changes across the lifespan is a central challenge for neuroscience, yet research on cognitive aging has remained largely disconnected from the theoretical and computational advances that now shape modern systems neuroscience. Over the past two decades, theoretical frameworks have transformed how we study cognition in young, healthy brains – providing principled tools to model latent decision states, neural dynamics, population codes, and inter-areal communication. In contrast, aging research has often relied on single-metric behavioral readouts, cross-sectional comparisons, and descriptive neural analyses, limiting our ability to explain fundamental differences in individual aging trajectories. This gap represents a missed opportunity: aging offers a powerful platform for testing theories of neural computation, stability, and flexibility under changing biological constraints. Here, we argue that closer integration between aging research and contemporary theoretical neuroscience can move the field beyond descriptive accounts toward more mechanistic explanations of decision-making across the lifespan. To this end, we outline how recent advances in behavioral quantification, latent state modeling, dynamical systems, encoding models, representational geometry, and recurrent neural networks offer a rich theoretical toolkit for neuroscientists studying decision-making across the lifespan.

**Introduction**
Research on the aging brain has traditionally centered on neurodegenerative diseases, particularly Alzheimer's disease and related dementias, yielding extensive insight into the pathological mechanisms of aging. However, far less is known about how neural circuits and computations necessary for decision-making change across the lifespan to support healthy aging. This is a missed opportunity to understand how common age-related biological constraints reshape decision-making, as well as cellular and circuit-level computations, in the absence of disease. Thus, normative aging represents both a major knowledge gap and a compelling opportunity to test modern theories of neural computation under changing biological conditions.

Despite substantial work, several fundamental questions in cognitive aging remain difficult to answer. First, aging is marked by considerable individual variability: people of similar chronological age and comparable post-mortem markers can show strikingly different cognitive function, yet most studies still rely on cross-sectional designs and across-subject averaging that often obscures such heterogeneity. Second, we lack a clear understanding of the longitudinal dynamics of aging: when age-related changes in the brain and behavior emerge, how rapidly they unfold, and how early alterations predict later outcomes. Third, it remains difficult to distinguish adaptive from maladaptive changes, as many neural and behavioral differences in older adults may reflect compensatory reorganization rather than decline, with not all deviations from a young adult brain being harmful. Addressing these challenges requires theoretical and computational approaches that can operate at the level of individuals, track cognitive states over time, and link behavior to complex neural dynamics. The approaches highlighted in this article, including recent advances in behavioral quantification, latent state modeling, dynamical systems, encoding models, representational geometry, and recurrent neural networks offer great advantages for studying decision-making across the lifespan.

**Advances in Quantifying Behavior**
Recent advances in behavioral tracking have substantially expanded how cognition and decision-making can be measured, moving beyond coarse single outcomes (e.g. left versus right or correct versus incorrect choices) or averaged summary statistics. High-resolution pose estimation methods (e.g. Lightning Pose [1]) and unsupervised movement-sequencing approaches such as motion sequencing (e.g. MoSeq [2]) enable continuous, multidimensional tracking of posture, kinematics, and action structure across time. Such richness is particularly interesting in the context of aging, where cognitive change is rarely isolated to a single domain but instead reflects coordinated shifts across perception, action, motivation, and control. Pose estimation methods such as DeepLabCut have already been applied to rodent decision-making tasks, revealing that movement trajectories encode latent decision variables including evidence strength, priors, and commitment [3], [4]. In parallel, unsupervised movement-sequencing approaches such as MoSeq demonstrate that behavior naturally decomposes into discrete latent states whose transitions are shaped by neural and neuromodulatory systems [5], [6]. Recent work has begun to explicitly connect these behavioral states to decision strategies [7], [8], providing a conceptual and methodological bridge between high-dimensional behavior and latent decision models. Applying these approaches to questions of cognitive aging can similarly uncover a wealth of information about nuanced changes in behavioral states.

For example, seminal work by Barnes and colleagues showed that cognitive aging is marked by instability of spatial representations rather than their loss. This landmark study found that aged rats formed accurate hippocampal place fields within sessions but failed to reliably reinstate the same maps across sessions in an unchanged environment [9] (see [10], [11] for more recent tests of this hypothesis in the dentate gyrus). These findings challenged deficit-based accounts and highlighted aging-related disruptions in the stabilization of internal representations over time. Until recently, however, it was difficult to link this neural instability to moment-to-moment behavioral state. Modern movement tracking and sequencing approaches now make it possible to jointly characterize neural and behavioral stability over time, enabling tests of whether representational drift coincides with shifts in navigation strategy, exploration, or decision commitment. Framing aging as a disruption in the stability or coupling of neural and behavioral states could extend Barnes' insights from descriptive remapping to a formal dynamical account of how internal models are maintained and reorganized with age.

Importantly, such high-dimensional behavioral features are naturally suited for integration within current state space modeling approaches, which can formalize latent cognitive or decision states and their transitions over time, providing a principled way to link continuous behavior to internal decision-making processes.

**Hidden Markov Models to Understand Latent Decision-Making and Neural States**
Hidden Markov Models (HMMs), a type of latent state space model, provide a framework for inferring latent ("hidden") states that evolve over time and govern observable behavior or neural activity. Conceptually, HMMs are well suited for capturing discrete cognitive or behavioral modes, such as strategies, internal states, or network dynamics, and the transitions between them, rather than assuming stationary processes (Fig. 1). In the behavioral domain, recent work has demonstrated the power of these approaches for modeling decision-making. For example, using a HMM to study reinforcement learning in young mice, Le and colleagues uncovered decision sub-strategies that go beyond traditional model-free vs model-based reinforcement-learning frameworks [12]. Furthermore, by combining the HMM with a generalized linear model (GLM) to characterize discrete states of sensory-guided decision-making, Ashwood and colleagues identified latent modes of engagement during decision-making which explain fluctuations in decision-making accuracy [13]. More recent work from Guo and colleagues, extends Ashwood's approach to study engagement during a reinforcement learning task to understand changes in decision-making in patients with major depressive disorder [14]. Finally, while not in the context of decision-making or cognition, work from Bedbrook and colleagues has begun to apply state space modeling to understand how behavioral states change across the lifespan in African killifish, a short-lived vertebrate model [15]. However, despite their clear relevance, these latent state behavioral models have rarely been applied to decision-making across the lifespan, representing a major gap in cognitive aging research.

Recently, HMMs have been applied to neural data from humans to study aging-related changes in brain dynamics. Early neuroimaging studies established that older adults often display broader or bilateral activation of decision-making brain regions relative to younger adults, termed *dedifferentiation*, alongside compensatory recruitment of additional regions in order to maintain cognitive performance [16], [17]. While these studies established that aging is associated with altered network recruitment and integration, they largely characterized *where* and *how much* neural activity differed between age groups, rather than *how neural states evolve over time*. HMMs extend this framework by explicitly modeling brain activity as transitions between latent states, enabling network-level activity to be understood as a dynamic process. When applied to neuroimaging data from humans, recent work has revealed that older adults exhibit altered expression, occupancy, and transition structure of latent neural states. Across MEG and fMRI studies, aging has been associated with prolonged dwell times in baseline and network states marked by high integration across brain regions, as well as reduced transitions between states, with systematic relationships between these latent state dynamics and cognitive performance, adding new detail to the idea of dedifferentiation [18], [19], [20]. While these studies are an excellent first step, they focus on characterizing dynamics of the brain in the resting state and will need to be extended to studying neural activity, as well as behavior, during decision-making across the lifespan. Furthermore, the inclusion of single-neuron data (rather than pooled population activity) will be essential, as critical, state-dependent changes in individual neurons are missed by measuring summed activity across neurons.

**Continuous and Switching Dynamics of Neural Activity**
Whereas HMMs capture cognitive computations as switches between discrete latent states, a linear dynamical system (LDS) models this as continuous latent states that evolve over time under linear dynamics. This makes it a natural tool to quantify how the aging brain may perform time-dependent computations differently, as implied by processing speed theories of cognitive aging [21], [22], [23]. In an aging context, LDS allows for explicit probing of dynamic variables. For example, decision-making dynamics in the aging brain may have longer timescales, leakier/noisier integration, or perturbed transitions between latent states.

Recurrent switching linear dynamical systems (rSLDS) build upon this framework by allowing linear dynamics to change abruptly when the system enters particular states, known as regimes, capturing both continuous and discrete shifts in population activity [13], [24], [25]. Work by Zoltowski and colleagues fit an rSLDS model to single-trial spiking data recorded from primate lateral intraparietal area (LIP) neurons during a random-dot motion discrimination task. Their model represents drift-diffusion-like evidence accumulation as a continuous latent integrator and choice commitment as a switch into a separate regime [26]. Use of simultaneous multi-region recordings could enable multi-population rSLDS, which infers both within and inter-area coupling between neurons that change in a dynamic, regime-dependent manner [27]. These dynamics functions enable explicit tests of whether aging alters state-dependent communication. Although this method is sparsely applied to aging datasets, one study utilizing Bayesian switching dynamical systems fit to fMRI data during a working memory task shows that older adults exhibit slower, more disorganized transitions between regimes [28]. Together, linear and switching dynamical systems offer a powerful yet largely untapped approach to quantifying how aging alters the temporal structure of neural computations during decision-making.

**Encoding Models to Understand Single Neuron Responses**
Encoding models seek to explain neural activity as a function of observable and latent variables, allowing researchers to test how task features, sensory inputs, movements, internal states, latent variables and even other neurons contribute to neural responses on a trial-by-trial basis (Fig. 2). Rather than treating neural variability as noise, these models ask which dimensions of behavior and cognition neurons represent, and how representations are organized within and across brain regions. Incorporating high-dimensional behavioral measurements alongside traditional decision variables has substantially expanded the power of this approach. For example, Musall and colleagues showed that much of the variance in cortical activity previously attributed to cognitive or sensory processes can be explained by detailed measures of movement, underscoring the importance of modeling behavior at fine temporal and spatial scales [29].

Recent extensions of encoding models further account for shared variability across neural populations by incorporating gain or coupling terms that capture coordinated fluctuations across neurons. These shared gain models allow neural responses to be modulated by latent factors reflecting internal states or shared inputs, enabling coupling both within regions [30], [31] and across different brain areas [32]. Closely related work has made inter-areal communication more explicit by formalizing how information is exchanged through low-dimensional communication subspaces [33], [34]. Notably, Srinath and colleagues demonstrated that changes in internal state, such as attention, can independently modulate local neural representations and the transmission of information between regions. These approaches are particularly powerful for aging research, where neural computations change across multiple levels, from single-neuron responses to network-level coordination, where aging could be modeled as a "state" change. By combining encoding models with rich measures of movement and decision-making, it becomes possible to determine whether age-related changes in neural activity reflect altered decision formation, memory/attentional engagement, movement, or broader alterations in network-level communication. What might once have been interpreted as generalized slowing can instead be decomposed into more specific underlying circuit mechanisms, offering a more precise and interpretable scaffold for studying decision-making across the lifespan.

**Representational Geometry to Capture High-Dimensional Features of Population Level Activity**
Representational geometry quantifies neural activity in terms of the structure with which information is represented at the population level [35], [36], [37]. By examining distances, angles, and manifolds of population activity in a high-dimensional space, this approach reveals how information is organized, separated, and generalized across task conditions (Fig. 3). Importantly, this framework is especially powerful for assessing what information a downstream brain region could plausibly extract, as geometric properties directly constrain the ease of linear or nonlinear readouts. In this way, representational geometry moves beyond the previously described approaches by describing how multiple variables can be represented in a neural population in a way that affords flexible decoding depending on the task at hand.

One recent example, from Bernardi and colleagues, used this approach, combined with electrophysiological recordings of the hippocampus (HP) and prefrontal cortex (PFC) in nonhuman primates (NHPs), to understand how the brain can flexibly classify different stimuli and generalize to novel contexts. This study exemplifies the unique strength of representational geometry by simultaneously assessing (1) how well neural activity in one task condition generalizes to a novel condition, termed cross-condition generalization and (2) how distinguishable neural activity in one condition is from all others, known as the shattering dimensionality. In a flexible decision-making task, the authors show that HP and PFC neurons in adult NHPs represent the context of their environment in an abstract manner that maintains both high cross-condition generalization and high shattering dimensionality [38]. These measures form the backbone of abstract and flexible decision-making, which is of particular relevance to aging researchers as cognitive flexibility is shown to decline with age [39], [40], [41]. Another recent study of representational geometry, by Lin and colleagues, extended this framework to more explicitly link geometric representations to the spiking of individual neurons. Using electrophysiological recordings of PFC from NHPs in a working memory task, they explicitly identify subpopulations of neurons that strongly contribute to the coding dimensions in geometric space in order to understand how their distinct *firing* properties might shape the *geometric* properties of overall population activity [42]. However, we currently lack information regarding how representational geometry changes with age. If applied to decision-making across the lifespan, these approaches could disambiguate whether and how aging impacts the ability of population activity to generalize across contexts, uniquely differentiate between similar stimuli, and/or support working memory during flexible decision-making.

**Modeling Circuit Mechanisms with Recurrent Neural Networks**
Recurrent neural networks (RNNs) model the evolution of population-level activity over time through recurrent interactions, providing a flexible framework for studying dynamical computations. Early work from Sussillo & Abbot shows that learning can sculpt the circuit properties of an RNN from chaotic spontaneous activity into stable, task-related activity *in silico* which mirrors circuit-level computations seen during learning *in vivo* [43]. Studies across primates, and later, rodents, apply this to context-dependent evidence integration tasks and successfully model selection and integration as emergent properties of recurrent computation (Fig. 4) [44], [45].

In contemporary systems neuroscience, trained RNNs are valuable because they can be optimized to reproduce behavior while also generating neural population dynamics comparable to neural recordings. Recent work has utilized RNNs to probe low-dimensional patterns and identify interpretable cognitive strategies through behavior in a variety of task-contexts [46], [47]. Moreover, this approach enables causal *in silico* perturbations of circuit parameters [48]. As such, they offer exciting opportunities to explore dysfunctional circuitry and maladaptive behavior, by isolating departures from normative learning dynamics and linking them to clinical phenotypes [49]. In aging, RNNs are attractive because many biological changes are naturally expressed as changes in network parameters such as reduced plasticity, altered connectivity, and increased noise. This provides the opportunity to explore and causally test which facets result in adaptive or maladaptive behaviors. Work from Ranjbar-Slamloo and colleagues demonstrate this in the mPFC of aging mice, whereby an RNN with age-dependent reductions in plasticity reproduced both functional dedifferentiation and shifts in network organization [50]. Importantly, by studying RNNs that successfully capture behavior and neural activity in aging individuals, we may uncover computational motifs that were absent in networks trained only on young individuals. By enabling direct manipulation of network parameters and dynamics, RNNs offer a unique and underutilized tool for causally testing how age-related biological changes give rise to adaptive or maladaptive decision-making strategies.

**Principles for Integrating Theory and Aging Research**
Taken together, these recent advances provide a concrete framework for the next phase of aging research in decision-making. First, studies should prioritize **within-individual and longitudinal designs**, leveraging latent-state and dynamical models to track how neural and behavioral computations evolve over time rather than relying solely on cross-sectional comparisons. Second, experiments should adopt **richer behavioral assays**, incorporating pose estimation, movement sequencing, fine-grained decision variables and model-based analysis [51] so that age-related neural changes can be more accurately linked to behavior. Third, analyses should increasingly incorporate **population-level and inter-areal models**, using encoding models, dynamical systems and representational geometry to determine how information is represented, transformed, and coordinated across circuits with age. Finally, computational models should be used not only to characterize data but to **generate testable predictions**, identifying which changes in dynamics, geometry, or connectivity are likely to be adaptive versus maladaptive and what interventions might be most effective. Viewing aging through the lens of these contemporary theoretical methods can move the field beyond descriptive accounts and towards mechanistic explanations that capture the complex, heterogenous effects of aging which reshape decision-making and the brain across the lifespan.

**Author credit statement:**
**Michael Ryan:** Conceptualization, Writing - Manuscript Preparation, Writing, Reviewing, Editing.
**Letizia Ye:** Writing – Manuscript Preparation, Writing, Reviewing, Editing.
**Anne Churchland:** Conceptualization, Writing - Manuscript Preparation, Writing, Reviewing, Editing.

**Declaration of competing interest:**
The authors declare that they have no known competing financial interests or personal relationships that could have appeared to influence the work reported in this paper.

**References:**

*Papers of particular interest, published within the period of review, have been highlighted as:*
*\* of special interest*
*\*\* of outstanding interest*

[1] D. Biderman *et al.*, "Lightning Pose: improved animal pose estimation via semi-supervised learning, Bayesian ensembling and cloud-native open-source tools," *Nat Methods*, vol. 21, no. 7, pp. 1316–1328, Jul. 2024, doi: 10.1038/s41592-024-02319-1.

[2] A. B. Wiltschko *et al.*, "Mapping Sub-Second Structure in Mouse Behavior," *Neuron*, vol. 88, no. 6, pp. 1121–1135, Dec. 2015, doi: 10.1016/j.neuron.2015.11.031.

[3] G. A. Kane, R. A. Senne, and B. B. Scott, "Rat movements reflect internal decision dynamics in an evidence accumulation task," *J Neurophysiol*, vol. 132, no. 5, pp. 1608–1620, Nov. 2024, doi: 10.1152/jn.00181.2024.

[4] M. Molano-Mazón *et al.*, "Rapid, systematic updating of movement by accumulated decision evidence," *Nat Commun*, vol. 15, no. 1, p. 10583, Dec. 2024, doi: 10.1038/s41467-024-53586-7.

[5] A. B. Wiltschko *et al.*, "Revealing the structure of pharmacobehavioral space through motion sequencing," *Nat Neurosci*, vol. 23, no. 11, pp. 1433–1443, Nov. 2020, doi: 10.1038/s41593-020-00706-3.

[6]\* J. E. Markowitz *et al.*, "Spontaneous behaviour is structured by reinforcement without explicit reward," *Nature*, vol. 614, no. 7946, pp. 108–117, Feb. 2023, doi: 10.1038/s41586-022-05611-2.

Using unsupervised behavioral segmentation, this study demonstrates how spontaneous actions can form structured sequences shaped by reinforcement, even in the absence of explicit reward.

[7]\*\* C. Yin *et al.*, "Spontaneous movements and their relationship to neural activity fluctuate with latent engagement states," *Neuron*, vol. 113, no. 18, pp. 3048-3063.e5, Sep. 2025, doi: 10.1016/j.neuron.2025.06.001.

In this study, the authors formally links fluctuations in engagement during a sensory-based decision-making task with the expression of task-dependent and task-independent movements, measured using DeepLabCut pose estimation in mice.

[8] D. Hulsey, K. Zumwalt, L. Mazzucato, D. A. McCormick, and S. Jaramillo, "Decision-making dynamics are predicted by arousal and uninstructed movements," *Cell Rep*, vol. 43, no. 2, p. 113709, Feb. 2024, doi: 10.1016/j.celrep.2024.113709.

[9] C. A. Barnes, M. S. Suster, J. Shen, and B. L. McNaughton, "Multistability of cognitive maps in the hippocampus of old rats," *Nature*, vol. 388, no. 6639, pp. 272–275, Jul. 1997, doi: 10.1038/40859.

[10] C. S. Herber, K. J. B. Pratt, J. M. Shea, S. A. Villeda, and L. M. Giocomo, "Spatial coding dysfunction and network instability in the aging medial entorhinal cortex," *Nat Commun*, vol. 16, no. 1, p. 8770, Oct. 2025, doi: 10.1038/s41467-025-63229-0.

[11] K. D. McDermott, M. A. Frechou, J. T. Jordan, S. S. Martin, and J. T. Gonçalves, "Delayed formation of neural representations of space in aged mice," *Aging Cell*, vol. 22, no. 9, p. e13924, Sep. 2023, doi: 10.1111/acel.13924.

[12] N. M. Le, M. Yildirim, Y. Wang, H. Sugihara, M. Jazayeri, and M. Sur, "Mixtures of strategies underlie rodent behavior during reversal learning," *PLoS Comput Biol*, vol. 19, no. 9, p. e1011430, Sep. 2023, doi: 10.1371/journal.pcbi.1011430.

[13] Z. C. Ashwood *et al.*, "Mice alternate between discrete strategies during perceptual decision-making," *Nat Neurosci*, vol. 25, no. 2, pp. 201–212, Feb. 2022, doi: 10.1038/s41593-021-01007-z.

[14] X. Guo, D. Zeng, and Y. Wang, "HMM for discovering decision-making dynamics using reinforcement learning experiments," *Biostatistics*, vol. 26, no. 1, p. kxae033, Dec. 2024, doi: 10.1093/biostatistics/kxae033.

[15] C. N. Bedbrook, R. D. Nath, L. Zhang, S. W. Linderman, A. Brunet, and K. Deisseroth, "Life-long behavioral screen reveals an architecture of vertebrate aging," Nov. 24, 2025, *bioRxiv*. doi: 10.1101/2025.11.21.688112.

[16] L. Aron, J. Zullo, and B. A. Yankner, "The adaptive aging brain," *Curr Opin Neurobiol*, vol. 72, pp. 91–100, Feb. 2022, doi: 10.1016/j.conb.2021.09.009.

[17] J. D. Koen and M. D. Rugg, "Neural Dedifferentiation in the Aging Brain," *Trends Cogn Sci*, vol. 23, no. 7, pp. 547–559, Jul. 2019, doi: 10.1016/j.tics.2019.04.012.

[18] R. Tibon, K. A. Tsvetanov, D. Price, D. Nesbitt, C. Can, and R. Henson, "Transient neural network dynamics in cognitive ageing," *Neurobiol Aging*, vol. 105, pp. 217–228, Sep. 2021, doi: 10.1016/j.neurobiolaging.2021.01.035.

[19]** M. Moretto, E. Silvestri, A. Zangrossi, M. Corbetta, and A. Bertoldo, "Unveiling whole-brain dynamics in normal aging through Hidden Markov Models," *Hum Brain Mapp*, vol. 43, no. 3, pp. 1129–1144, Feb. 2022, doi: 10.1002/hbm.25714.

This study demonstrates the power of Hidden Markov Models in aging research to identify age-related changes in whole-brain neural state dynamics from fMRI, revealing altered state occupancy and transition structure in aging humans.

[20] K. Chen *et al.*, "Hidden Markov Modeling Reveals Prolonged 'Baseline' State and Shortened Antagonistic State across the Adult Lifespan," *Cereb Cortex*, vol. 32, no. 2, pp. 439–453, Jan. 2022, doi: 10.1093/cercor/bhab220.

[21] T. A. Salthouse and W. Lichty, "Tests of the neural noise hypothesis of age-related cognitive change," *J Gerontol*, vol. 40, no. 4, pp. 443–450, Jul. 1985, doi: 10.1093/geronj/40.4.443.

[22] T. A. Salthouse, "The processing-speed theory of adult age differences in cognition," *Psychol Rev*, vol. 103, no. 3, pp. 403–428, Jul. 1996, doi: 10.1037/0033-295x.103.3.403.

[23] R. Ratcliff, A. Thapar, and G. McKoon, "Aging and individual differences in rapid two-choice decisions," *Psychon Bull Rev*, vol. 13, no. 4, pp. 626–635, Aug. 2006, doi: 10.3758/bf03193973.

[24] S. W. Linderman, M. J. Johnson, M. A. Wilson, and Z. Chen, "A Bayesian nonparametric approach for uncovering rat hippocampal population codes during spatial navigation," *J Neurosci Methods*, vol. 263, pp. 36–47, Apr. 2016, doi: 10.1016/j.jneumeth.2016.01.022.

[25] S. W. Linderman and S. J. Gershman, "Using computational theory to constrain statistical models of neural data," *Curr Opin Neurobiol*, vol. 46, pp. 14–24, Oct. 2017, doi: 10.1016/j.conb.2017.06.004.

[26] D. M. Zoltowski, K. W. Latimer, J. L. Yates, A. C. Huk, and J. W. Pillow, "Discrete Stepping and Nonlinear Ramping Dynamics Underlie Spiking Responses of LIP Neurons during Decision-Making," *Neuron*, vol. 102, no. 6, pp. 1249-1258.e10, Jun. 2019, doi: 10.1016/j.neuron.2019.04.031.

[27] J. Glaser, M. Whiteway, J. P. Cunningham, L. Paninski, and S. Linderman, "Recurrent Switching Dynamical Systems Models for Multiple Interacting Neural Populations," in *Advances in Neural Information Processing Systems*, Curran Associates, Inc., 2020, pp. 14867–14878. Accessed: Jan. 24, 2026. [Online]. Available: https://proceedings.neurips.cc/paper/2020/hash/aa1f5f73327ba40d47ebce155e785aaf-Abstract.html

[28]** B. Lee *et al.*, "Latent brain state dynamics and cognitive flexibility in older adults," *Prog Neurobiol*, vol. 208, p. 102180, Jan. 2022, doi: 10.1016/j.pneurobio.2021.102180.

This study on aging leverages a linear dynamical systems model to characterize continuous latent brain state dynamics during working memory and found that aging was associated with disrupted and slower state transitions in humans.

[29] S. Musall, M. T. Kaufman, A. L. Juavinett, S. Gluf, and A. K. Churchland, "Single-trial neural dynamics are dominated by richly varied movements," *Nat Neurosci*, vol. 22, no. 10, pp. 1677–1686, Oct. 2019, doi: 10.1038/s41593-019-0502-4.

[30] E. Hart and A. C. Huk, "Recurrent circuit dynamics underlie persistent activity in the macaque frontoparietal network," *Elife*, vol. 9, p. e52460, May 2020, doi: 10.7554/eLife.52460.

[31]** C. Haimerl, D. A. Ruff, M. R. Cohen, C. Savin, and E. P. Simoncelli, "Targeted V1 comodulation supports task-adaptive sensory decisions," *Nat Commun*, vol. 14, no. 1, p. 7879, Nov. 2023, doi: 10.1038/s41467-023-43432-7.

This work employs an encoding model with a shared gain modulation, demonstrating that latent, population-wide modulatory signals dynamically shape sensory responses in V1 during sensory discrimination in nonhuman primates.

[32] N. C. Rabinowitz, R. L. Goris, M. Cohen, and E. P. Simoncelli, "Attention stabilizes the shared gain of V4 populations," *Elife*, vol. 4, p. e08998, Nov. 2015, doi: 10.7554/eLife.08998.

[33] J. D. Semedo, A. Zandvakili, C. K. Machens, B. M. Yu, and A. Kohn, "Cortical Areas Interact through a Communication Subspace," *Neuron*, vol. 102, no. 1, pp. 249-259.e4, Apr. 2019, doi: 10.1016/j.neuron.2019.01.026.

[34] R. Srinath, D. A. Ruff, and M. R. Cohen, "Attention improves information flow between neuronal populations without changing the communication subspace," *Curr Biol*, vol. 31, no. 23, pp. 5299-5313.e4, Dec. 2021, doi: 10.1016/j.cub.2021.09.076.

[35] M. Rigotti *et al.*, "The importance of mixed selectivity in complex cognitive tasks," *Nature*, vol. 497, no. 7451, pp. 585–590, May 2013, doi: 10.1038/nature12160.

[36]  U. Cohen, S. Chung, D. D. Lee, and H. Sompolinsky, "Separability and geometry of object manifolds in deep neural networks," *Nat Commun*, vol. 11, no. 1, p. 746, Feb. 2020, doi: 10.1038/s41467-020-14578-5.

[37]  L. Posani, S. Wang, S. P. Muscinelli, L. Paninski, and S. Fusi, "Rarely categorical, always high-dimensional: how the neural code changes along the cortical hierarchy," *bioRxiv*, p. 2024.11.15.623878, Feb. 2025, doi: 10.1101/2024.11.15.623878.

[38]**  S. Bernardi, M. K. Benna, M. Rigotti, J. Munuera, S. Fusi, and C. D. Salzman, "The Geometry of Abstraction in the Hippocampus and Prefrontal Cortex," *Cell*, vol. 183, no. 4, pp. 954-967.e21, Nov. 2020, doi: 10.1016/j.cell.2020.09.031.

This is an exceptional study on representational geometry, offering clear intuition regarding how geometric representations in population activity can impact the representation of an abstract rule during flexible decision-making in nonhuman primates.

[39]  B. Eppinger, A. Ruel, and F. Bolenz, "Diminished State Space Theory of Human Aging," *Perspect Psychol Sci*, vol. 20, no. 2, pp. 325–339, Mar. 2025, doi: 10.1177/17456916231204811.

[40]  J. Strough and W. Bruine de Bruin, "Decision Making Across Adulthood," Dec. 01, 2020, *Social Science Research Network, Rochester, NY*: 3750568. doi: 10.1146/annurev-devpsych-051120-010038.

[41]  E. P. Sparrow and J. Spaniol, "Age-Related Changes in Decision Making," *Curr Behav Neurosci Rep*, vol. 3, no. 4, pp. 285–292, Dec. 2016, doi: 10.1007/s40473-016-0091-4.

[42]* X.-X. Lin, A. Nieder, and S. N. Jacob, "The neuronal implementation of representational geometry in primate prefrontal cortex," *Sci Adv*, vol. 9, no. 50, p. eadh8685, Dec. 2023, doi: 10.1126/sciadv.adh8685.

A recent paper which links population-level representational geometry in prefrontal cortex to the firing properties of individual neurons by identifying subpopulations that strongly contribute to specific coding dimensions.

[43]  D. Sussillo and L. F. Abbott, "Generating coherent patterns of activity from chaotic neural networks," *Neuron*, vol. 63, no. 4, pp. 544–557, Aug. 2009, doi: 10.1016/j.neuron.2009.07.018.

[44]* V. Mante, D. Sussillo, K. V. Shenoy, and W. T. Newsome, "Context-dependent computation by recurrent dynamics in prefrontal cortex," *Nature*, vol. 503, no. 7474, pp. 78–84, Nov. 2013, doi: 10.1038/nature12742.

Seminal work which trained recurrent neural networks to perform a context-dependent evidence integration task, showing that selection of relevant sensory information emerges from recurrent population dynamics rather than explicit gating.

[45]  M. Pagan *et al.*, "Individual variability of neural computations underlying flexible decisions," *Nature*, vol. 639, no. 8054, pp. 421–429, Mar. 2025, doi: 10.1038/s41586-024-08433-6.

[46]  L. Ji-An, M. K. Benna, and M. G. Mattar, "Discovering cognitive strategies with tiny recurrent neural networks," *Nature*, vol. 644, no. 8078, pp. 993–1001, Aug. 2025, doi: 10.1038/s41586-025-09142-4.

[47] K. Miller, M. Eckstein, M. Botvinick, and Z. Kurth-Nelson, "Cognitive Model Discovery via Disentangled RNNs," *Advances in Neural Information Processing Systems*, vol. 36, pp. 61377–61394, Dec. 2023.

[48] J. P. Roach, A. K. Churchland, and T. A. Engel, "Choice selective inhibition drives stability and competition in decision circuits," *Nat Commun*, vol. 14, no. 1, p. 147, Jan. 2023, doi: 10.1038/s41467-023-35822-8.

[49] A. Dezfouli, K. Griffiths, F. Ramos, P. Dayan, and B. W. Balleine, "Models that learn how humans learn: The case of decision-making and its disorders," *PLoS Comput Biol*, vol. 15, no. 6, p. e1006903, Jun. 2019, doi: 10.1371/journal.pcbi.1006903.

[50] Y. Ranjbar-Slamloo, H. R. Chong, and T. Kamigaki, "Aging disrupts the link between network centrality and functional properties of prefrontal neurons during memory-guided behavior," *Commun Biol*, vol. 8, no. 1, p. 62, Jan. 2025, doi: 10.1038/s42003-025-07498-x.

By training recurrent neural networks on a memory-guided task and systematically altering plasticity parameters, this study captures key features of age-related circuit reorganization in prefrontal cortex in mice.

[51] M. Carandini and A. K. Churchland, "Probing perceptual decisions in rodents," *Nat Neurosci*, vol. 16, no. 7, pp. 824–831, Jul. 2013, doi: 10.1038/nn.3410.

[52] B. Bagi, M. Brecht, and J. I. Sanguinetti-Scheck, "Unsupervised discovery of behaviorally relevant brain states in rats playing hide-and-seek," *Curr Biol*, vol. 32, no. 12, pp. 2640-2653.e4, Jun. 2022, doi: 10.1016/j.cub.2022.04.068.

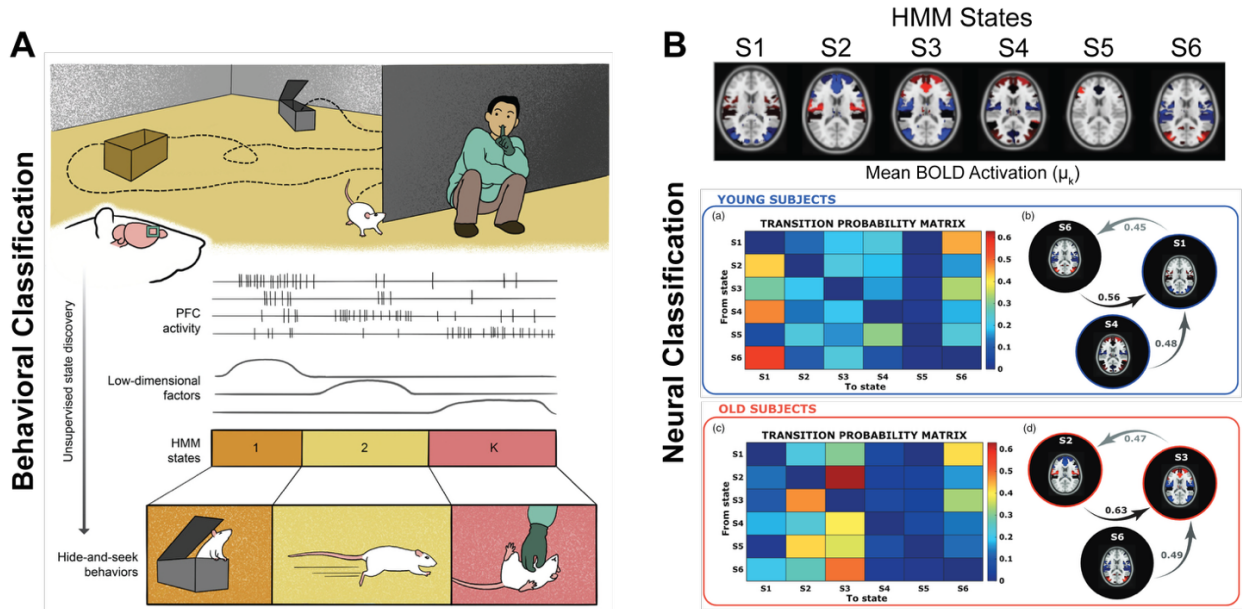

**Fig 1. Hidden Markov Models to Understand Latent Decision-Making and Neural States**
**A. See Bagi et al., 2022.** Using a Hidden Markov Model (HMM) to classify neural activity recorded in prefrontal cortex (PFC) from young rats performing a game of hide-and-seek, Bagi and colleagues discover latent neural states that map onto discrete phases of the animal's spontaneous behavior [52]. **B. Adapted from Moretto et al., 2022.** Training a HMM on resting state fMRI data from young and old subjects, Moretto and colleagues found 6 distinct latent states in neural activity (**top**). When comparing the probability of neural activity transitioning from one state to the next, they found that older subjects showed increased occupancy and transitions into neural states characterized by high integration (e.g. S2,S3) compared to younger subjects whose neural states transitioned largely between highly segregated (e.g. S1,S6) neural states (**bottom**) [19].

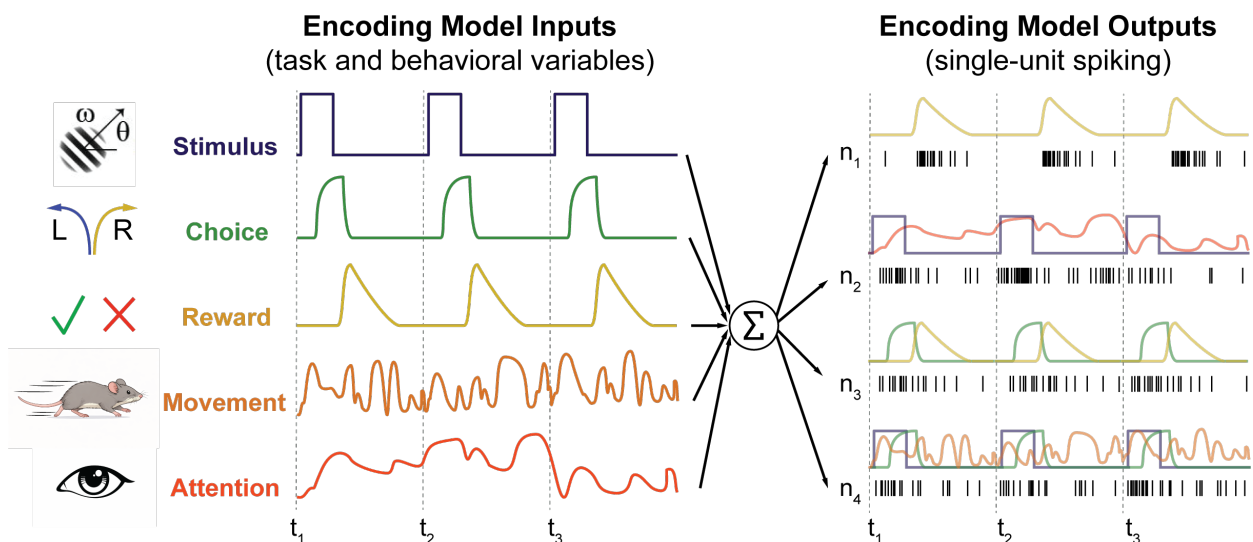

**Fig 2. Encoding Models to Understand Single Neuron Responses**
**(Left)** Schematic of the behavioral inputs to an encoding model, with time-varying signals for both task-related (e.g. stimulus, choice, and reward) and spontaneous (e.g. movement and attention) behaviors depicted across three trials (t). **(Right)** Schematized raster plots for four neurons (n), whose activity is best predicted by the encoding model with high weights assigned to different behavioral variables. Using this approach, trial-to-trial differences in firing rates can be better explained by the mixed selectivity of neurons for different combinations of task-related and spontaneous behaviors.

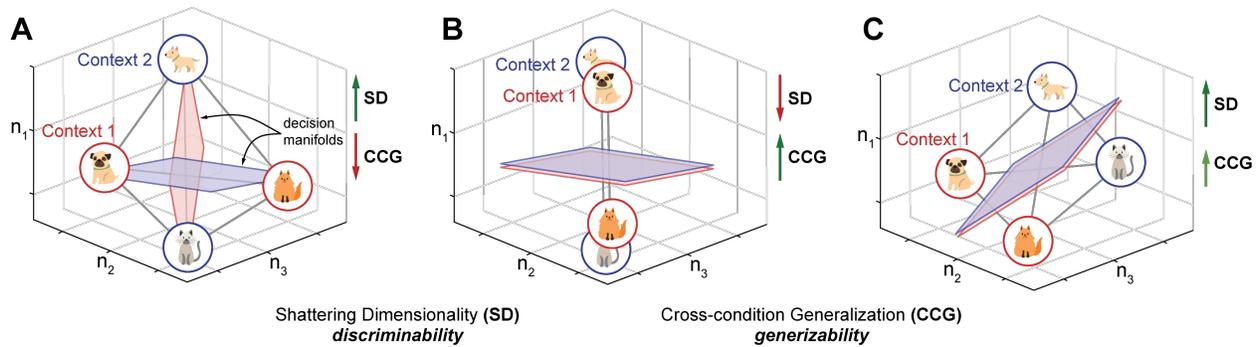

**Fig 3. Representational Geometry to Capture High-Dimensional Features of Population Level Activity.**
**Adapted from Bernardi et al., 2020 [38].** Representational geometry describes how abstract decision rules can be flexibly represented in the brain. Here, we can plot the activity of three simultaneously recorded neurons (n) in a 3D population space. Now imagine a task where a participant is trained to differentially categorize two pictures in a given context, such as the cat and dog in Context 1 (red). What happens when the participant is now asked to categorize two novel pictures they have not encountered before, such as Context 2 (blue), using the same decision rule? As it turns out, the answer depends on the high dimensional geometric relationship of representations between the population of neurons, known as representational geometry. **(A)** In this geometric representation, the neurons maximally discriminate between all the stimuli presented in both contexts, known as shattering dimensionality (SD). In high dimensional space, any stimulus can be easily separated from all other stimuli. However, this geometry makes generalization between contexts, known as cross-condition generalization, impossible. A decision manifold used to build an abstract rule classifying a dog from a cat in Context 1 (blue manifold) is orthogonal to the decision manifold in Context 2 (red manifold). **(B)** Alternatively, neurons could maximize generalization by having the neural population represent dogs similarly and cat similarly, regardless of context. While these decision manifolds are now well aligned to generalize effortlessly to novel stimuli, we can no longer discriminate between different stimuli of the same type (e.g. between dogs). **(C)** Finally, we have a geometry that attempts to maximize both discriminability and generalization. Here, while each stimulus is still discriminable from all other stimuli, by rotating their representations in population space, their decision manifolds can also be closely aligned allowing for robust generalization. By recording from neurons in the prefrontal cortex and hippocampus, Bernardi and colleagues show that this is precisely how population activity in non-human primates is organized during a flexible decision-making task.

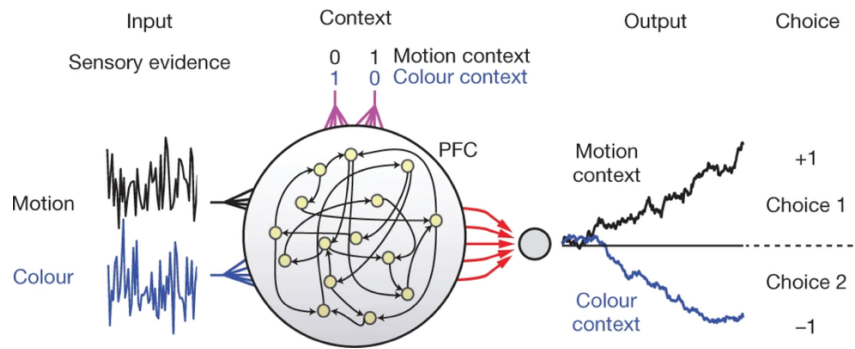

**Fig 4. Modeling Circuit Mechanisms with Recurrent Neural Networks**
**See Mante et al., 2013.** Using a recurrent neural network (RNN), Mante and colleagues modeled prefrontal cortical (PFC) activity in non-human primates performing a flexible visual decision-making task that required context-dependent integration of noisy sensory inputs. Animals were trained to make a saccadic eye movement to a choice target based on either the motion or color of a random-dot stimulus, depending on a concurrent contextual cue indicating which feature was relevant on a given trial. To test the hypothesis that irrelevant sensory information is filtered prior to integration, the authors modeled PFC as a fully recurrent network of nonlinear units receiving independent motion, color, and contextual inputs. The network was then trained to generate a binary choice output (e.g. +1 for choice 1 and -1 for choice 2) through a linear downstream readout. Rather than observing early gating of irrelevant inputs, the model revealed that both relevant and irrelevant sensory signals are initially mixed within single PFC neurons, with selection emerging late through the same recurrent circuitry that integrates evidence toward the choice. This late selection is possible because task variables and upcoming choices are separable at the population level, despite substantial mixing in single-neuron activity.